\documentclass[aps,prd,superscriptaddress,twoside,twocolumn,nofootinbib,10pt,
showpacs,floatfix]{revtex4-2}
\usepackage{graphicx,lineno,color,xcolor}
\usepackage{hyperref,url}
\usepackage[title]{appendix}
\usepackage{amsmath}
\usepackage{amssymb}

\newcommand{\dd}{{\rm d}}
\newcommand{\aap}{Astro.~Astrophys.}    \newcommand{\mnras}{Mon.~Not.~R.~Astron.~Soc.}
\newcommand{\physrep}{Phys.~Rep.}       \newcommand{\ijmpe}{Int.~J.~Mod.~Phys.~E}
\begin{document}
\title{The continuous wavelet derived by smoothing function and its application in cosmology}
\author{Yun Wang}
\affiliation{College of Physics, Jilin University, Changchun 130012, China}
\author{Ping He}\email{hep@jlu.edu.cn}
\affiliation{College of Physics, Jilin University, Changchun 130012, China}
\affiliation{Center for High Energy Physics, Peking University, Beijing 100871, China}
\date{\today}
\begin{abstract}
The wavelet analysis technique is a powerful tool and is widely used in broad disciplines of engineering, technology, and sciences. In this work, we present a novel scheme of constructing continuous wavelet functions, in which the wavelet functions are obtained by taking the first derivative of smoothing functions with respect to the scale parameter. Due to this wavelet constructing scheme, the inverse transforms are only one-dimensional integrations with respect to the scale parameter, and hence the continuous wavelet transforms constructed in this way are more ready to use than the usual scheme. We then apply the Gaussian-derived wavelet constructed by our scheme to computations of the density power spectrum for dark matter, the velocity power spectrum and the kinetic energy spectrum for baryonic fluid. These computations exhibit the convenience and strength of the continuous wavelet transforms. The transforms are very easy to perform, and we believe that the simplicity of our wavelet scheme will make continuous wavelet transforms very useful in practice.
{\vskip 10pt}
\noindent {\bf Key words:} wavelet analysis, intergalactic medium, large-scale structure of Universe
\end{abstract}

\maketitle

\section{Introduction}
\label{sec:intro}

It is well known that the Fourier transform has many shortcomings, of which the most is that it cannot simultaneously provide information about the scale and position of the signal \cite{Fang1997}. To overcome the drawbacks of the Fourier transform, the wavelet analysis technique was invented and has proven a powerful tool, and is now widely used in areas such as signal processing, image and data compactness \cite{Daubechies1992, Chui1992}. The wavelet analysis has also been extensively applied to astrophysics and cosmology for more than three decades. The continuous wavelet transform (CWT) is applied to, to name a few, the analysis of the large-scale structures of the Universe \cite{Slezak1990, Escalera1992a, Escalera1992b, Martinez1993, Fujiwara1996}, the detection of point sources or patterns of astronomical objects \cite{Cayon2000, Vielva2003, Gonzalez-Nuevo2006, Batista2011, Mertens2015, Baluev2018}, the detection of the non-Gaussianity in the CMB maps \cite{Barreiro2000, Cayon2001, Aghanim2003}, the foreground or noise subtraction in the CMB maps or sky surveys \cite{Sanz1999, Hansen2006, Gu2013}, the cosmological $N$-body simulation with high performances \cite{Romeo2003, Romeo2004}. Besides the CWT, the discrete wavelet transform (DWT) is also used in studies of cosmological large-scale structures \cite{Pando1996, Pando1998a, Pando1998b, Fang2000}.

As mentioned above, there are two kinds of wavelet transforms, namely, the CWT and the DWT. The CWT provides an overcomplete representation of a signal by varying continuously the scale and translation parameter of the wavelets. With $a$ and $b$ as scale and translation parameter respectively, the CWT of function $f(x)$ is defined as
\begin{equation}
\label{eq:cwt}
WT_f(a,b) = \frac{1}{\sqrt{|a|}}\int^{\infty}_{-\infty} f(x)\bar{\psi}(\frac{x-b}{a})\dd x,
\end{equation}
where $\psi(x)$ is called the mother wavelet, and the over-bar indicates the complex conjugate. $\psi(x)$ should meet the square-integrable condition,
\begin{equation}
\label{eq:sicond}
\int_{-\infty}^{\infty}|\psi(x)|^2 \dd x = \frac{1}{2\pi} \int_{-\infty}^{\infty}|\hat{\psi}(k)|^2 \dd k < \infty,
\end{equation}
in which $\hat{\psi}(k)$ is the Fourier transform of $\psi(x)$, and we use Parseval's theorem. The inverse transform of Eq.~(\ref{eq:cwt}) is
\begin{equation}
\label{eq:icwt}
f(x) = C^{-1}_{\psi}\int^{\infty}_{-\infty}\int^{\infty}_{-\infty} WT_f(a,b) \frac{1}{\sqrt{|a|}}\psi(\frac{x-b}{a})\frac{\dd a \dd b}{a^2},
\end{equation}
with
\begin{equation}
\label{eq:cpsi}
C_{\psi}=\int^{\infty}_{-\infty}\frac{|\hat{\psi}(k)|^2}{|k|}\dd k,
\end{equation}
where $C_{\psi}$ should satisfy the so-called admissibility condition $0<C_{\psi}<\infty$, which suggests that $\hat{\psi}(0)=0$.

The advantage of the CWTs is that they can provide arbitrary localization in the scale and position due to their continuity. On the contrary, the DWTs do not vary continuously in the scale and translation parameter, and for a function $f(x)$, its DWT usually takes the dyadic form as
\begin{equation}
\label{eq:dwt}
\tilde{\epsilon}_{j,l} = \sqrt{\frac{2^j}{L}} \int_{-\infty}^{\infty}f(x)\bar{\psi}(2^{j}x/L-l)\dd x,
\end{equation}
and $f(x)$ can be expressed by the wavelet coefficients $\tilde{\epsilon}_{j,l}$ from the inverse transform as
\begin{equation}
\label{eq:idwt}
f(x) = \sqrt{\frac{2^j}{L}} \sum^{\infty}_{j=0} \sum^{\infty}_{l=-\infty} \tilde{\epsilon}_{j,l} \psi(2^jx/L - l).
\end{equation}
For a comprehensive understanding of the CWT and the DWT, we refer the reader to Refs.~\cite{Daubechies1992, Chui1992}.

The advantage of the DWTs constructed in this way is that they can provide a set of complete and orthonormal bases $\psi_{j,l}(x) = (2^j/L)^{1/2}\psi(2^jx/L - l)$ for the wavelet expansion.

The disadvantage of the usual CWTs is that the inverse transform is a two-dimensional integration as Eq.~(\ref{eq:icwt}), which may be computationally very complex and difficult; while DWTs are awkward to use since both spatial translation and scale dilation have to be dyadically adopted, that is, the dilation or translation cannot be arbitrarily performed. As a result, the expressions of the DWT power spectrum are complicated and cumbersome \cite{Fang2000, Pando1998c, Yang2001}.

In this work, we present a novel scheme of constructing continuous wavelet functions, in which the wavelet functions are obtained by taking the first derivative of smoothing functions with respect to the scale parameter, and the inverse transforms are only one-dimensional (1D) integrations, and hence the CWTs constructed in this way are more ready to use than the usual forms. The paper is organized as follows. We outline the basic theoretical framework in Section~\ref{sec:basis}, and give the fast algorithm of our wavelet transform in Section~\ref{sec:fast}. We present some simple applications in cosmology of the wavelet technique in Section~\ref{sec:app}. In Sections~\ref{sec:disc}, we give some discussions of the results, and present the summary and conclusions in Section~\ref{sec:conc}.

\begin{figure}
\includegraphics[width=1.0\linewidth]{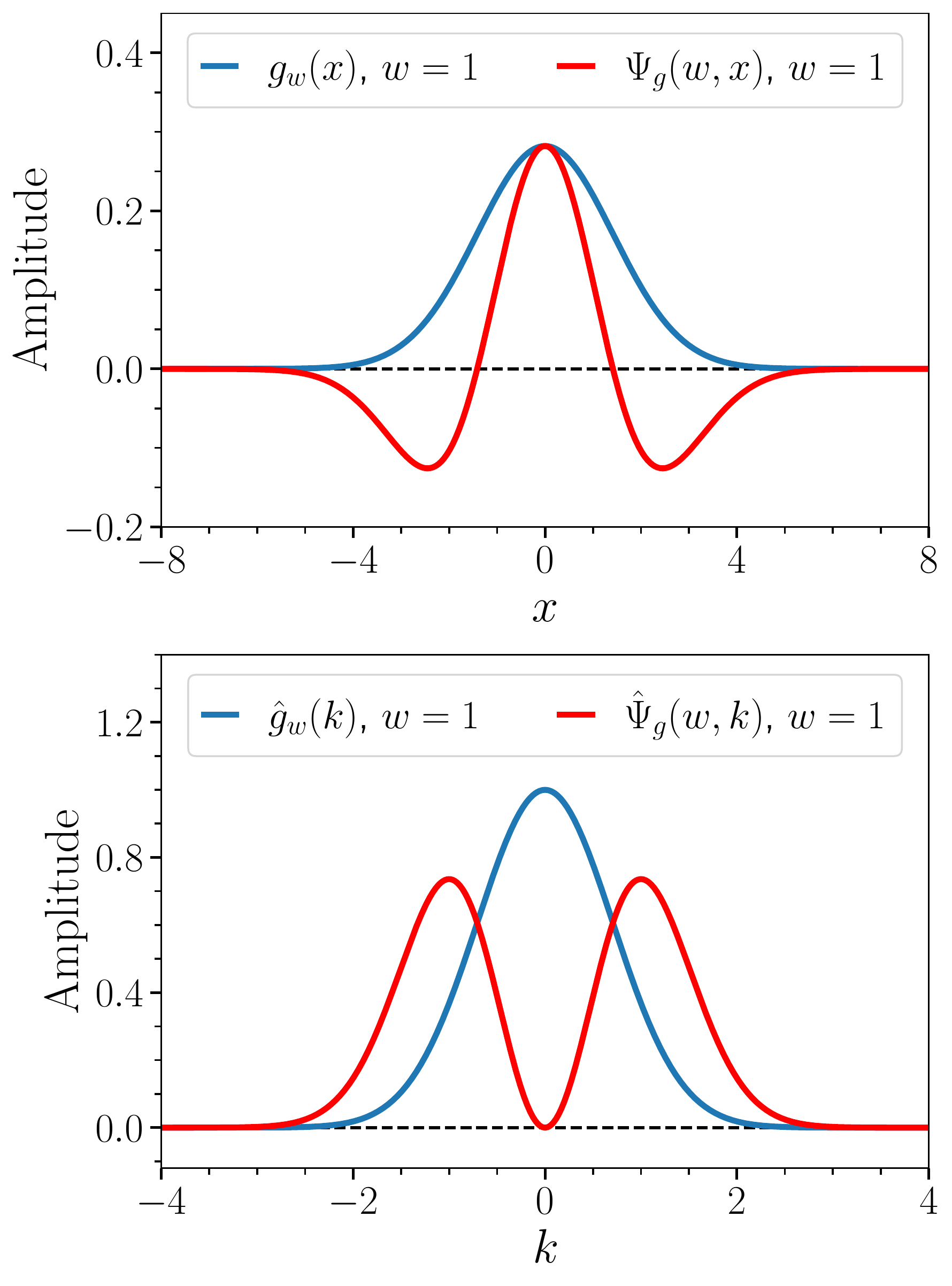}
\caption{The upper panel is for the Gaussian and the wavelet function; the lower panel is for their corresponding Fourier transforms. We see that the Gaussian is a low-pass filter, while the wavelet is a band-pass filter.}
\label{fig:wvlet}
\end{figure}

\section{Basic theoretical framework}
\label{sec:basis}

We describe the process of constructing the CWTs in the following. As a comparison, the detailed derivation of the traditional CWT and the inverse transform are given explicitly in Appendix~\ref{appendix:traditional_cwt}.

Generally, a smoothing function $s_w(x)$ with the scale parameter $w$ is an even function of $x$, i.e. $s_w(x)=s_w(-x)$, and should satisfy the normalization condition
\begin{equation}
\label{eq:norm}
\int_{-\infty}^{+\infty}s_w(x)\dd x=1.
\end{equation}
Instead of a general presentation, however, we proceed with a concrete example, i.e. the Gaussian function $g_{\sigma}(x)=e^{-x^2/2\sigma^2}/ (\sqrt{2\pi}\sigma)$, with $\sigma>0$. Letting $\sigma = \sqrt{2}/w$, we have
\begin{equation}
\label{eq:gaussian}
g_w(x)=\frac{w}{2\sqrt{\pi}}e^{-\frac{1}{4}w^2x^2},
\end{equation}
which satisfies the normalization condition Eq.~(\ref{eq:norm}) and is surely a smoothing function. In the following, we call $w$ the scale parameter. When $w\rightarrow\infty$, $g_w(x)$ degenerates to Dirac $\delta$ function, i.e. $g_w(x)\rightarrow\delta_D(x)$. The Fourier transform of $g_w(x)$ is
\begin{equation}
\label{eq:gaussft}
\hat{g}_w(k)=e^{-\frac{k^2}{w^2}}.
\end{equation}
We see that $\hat{g}_w(0)=1$, consistent with the normalization condition Eq.~(\ref{eq:norm}). Additionally, the reason for choosing $w$ as $w=\sqrt{2}/\sigma$ is that $w$ can one-to-one correspond to $k$, as can be seen from Eq.~(\ref{eq:gaussft}). In Fig.~\ref{fig:wvlet}, we show the curves of $g_w(x)$ and $\hat{g}_w(k)$.

For a random field, such as the density contrast $\delta(x)$ of the cosmological density, the smoothed field $\delta_g(w,x)$ under the scale parameter $w$ can be obtained via the convolution with the smoothing function $g_w(x)$ as
\begin{equation}
\label{eq:smt}
\delta_g(w,x)=\int_{-\infty}^{+\infty}\delta(u)g_w(x-u)\dd u.
\end{equation}
When $w\rightarrow\infty$, then $g_w(x)\rightarrow\delta_D(x)$, and hence $\delta_g(w,x)\rightarrow \delta(x)$. Taking the first derivative of both sides of Eq.~(\ref{eq:smt}) with respect to $w$, we have
\begin{eqnarray}
\label{eq:gaussdev}
W_{\delta}(w,x) & \equiv &\frac{\partial{\delta_g(w,x)}}{\partial{w}} = \int_{-\infty}^{+\infty} \delta(u) \frac{\partial{g_w(x-u)}}{\partial{w}} \dd u \nonumber \\
& = & \int_{-\infty}^{+\infty} \delta(u) \Psi_g(w,x-u)\dd u,
\end{eqnarray}
in which we define $\Psi_g$ as
\begin{equation}
\label{eq:gausswv}
\Psi_g(w,x)\equiv\frac{\partial{g_w(x)}}{\partial{w}}=\frac{1}{4\sqrt{\pi}}(2-w^2x^2) e^{-\frac{1}{4} w^2x^2},
\end{equation}
which is the {\em Gaussian-derived wavelet} that we call, and is nothing but the 1D Mexican hat wavelet function. The Fourier transform of Eq.~(\ref{eq:gausswv}) is
\begin{equation}
\label{eq:gausswvft}
\hat{\Psi}_g(w,k) = \frac{2k^2}{w^3} e^{-\frac{k^2}{w^2}}.
\end{equation}
We also plot $\Psi_g(w,x)$ and $\hat{\Psi}_g(w,k)$ in Fig.~\ref{fig:wvlet}, from which we see that $\hat{g}_w(k)$ is a low-pass filter, while $\hat{\Psi}_g(w,k)$ is a band-pass filter. Note that $\hat{g}_w(k)$ and $\hat{\Psi}_g(w,k)$ are related by
\begin{equation}
\label{eq:wvft2}
\hat{\Psi}_g(w,k)=\frac{\partial{\hat{g}_w(k)}}{\partial{w}},
\end{equation}
which is consistent with the definition of $\Psi_g(w,x)$ in Eq.~(\ref{eq:gausswv}). Integrating Eq.~(\ref{eq:gaussdev}) with respect to $w$, we have
\begin{equation}
\label{eq:invwv1}
\delta_g(w,x) = \delta_g(0,x) + \int^{w}_{0}W_{\delta}(w',x)\dd w',
\end{equation}
where $\delta_g(0,x)$ is an integration constant. For cosmologically interesting objects, however, such as the density contrast field or the peculiar velocity field, the whole-cosmos averaged quantities should be usually vanishing, i.e. $\delta_g(0,x)=0$, or at least a constant that is independent of spatial positions, and hence we can safely neglect this term in the future. Notice that $\delta(x) = \delta_g(\infty,x)$, we obtain
\begin{equation}
\label{eq:invwv2}
\delta(x) = \int^{\infty}_{0}W_{\delta}(w,x)\dd w,
\end{equation}
which is just the inverse transform of the wavelet transform Eq.~(\ref{eq:gaussdev}). Compared with the usual inverse wavelet transform Eq.~(\ref{eq:icwt}), we see that our inverse transform is only 1D integration, which is much easier to manipulate than Eq.~(\ref{eq:icwt}).

\begin{figure}
\includegraphics[width=1.0\linewidth]{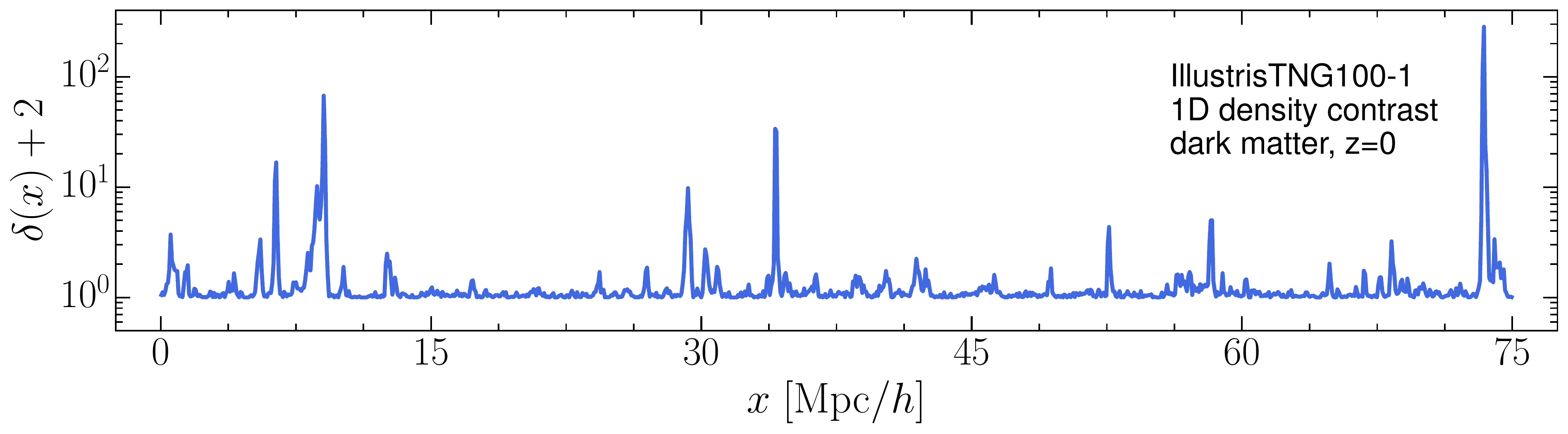}
\includegraphics[width=1.0\linewidth]{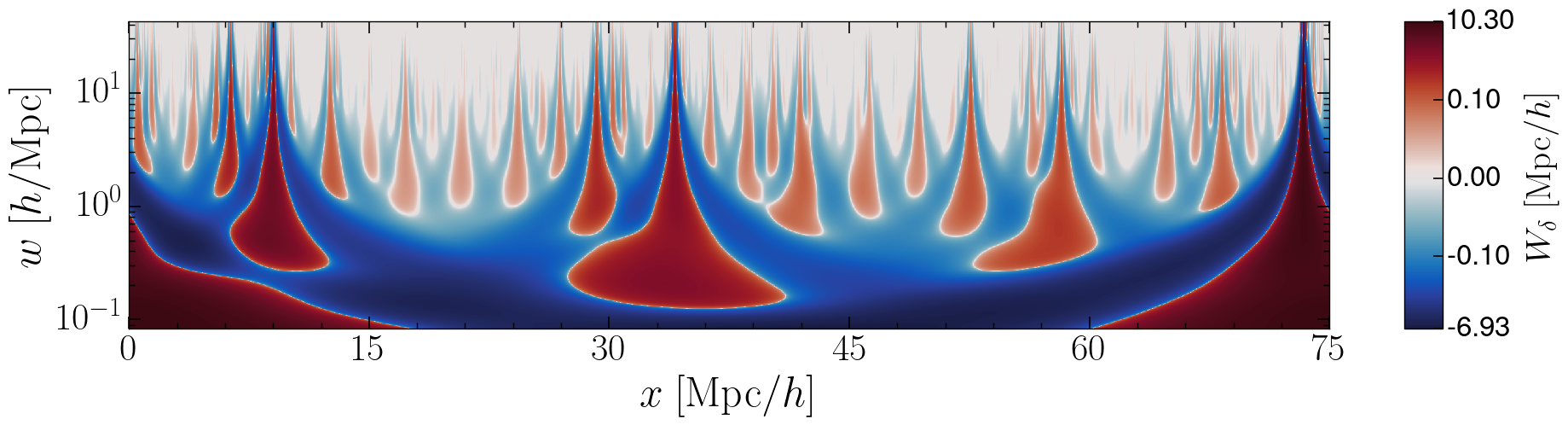}
\caption{Upper panel: the 1D density contrast field of dark matter. Lower panel: the corresponding wavelet scalogram of the 1D density contrast field. The data is taken from the IllustrisTNG simulation.}
\label{fig:denscalogram}
\end{figure}

\section{Fast algorithm of wavelet transform}
\label{sec:fast}

Due to the wavelet transform pair Eqs.~(\ref{eq:gaussdev}) and (\ref{eq:invwv2}), $W_{\delta}(w,x)$ is equivalent to the original field $\delta(x)$, and hence we can use $W_{\delta}(w,x)$ for further studies instead of $\delta(x)$. Given a density contrast (or other) field $\delta(x)$, we can calculate its wavelet transform directly using Eq.~(\ref{eq:gaussdev}). However, there exists a fast algorithm based on the technique of Fast Fourier Transform (FFT). From Eq.~(\ref{eq:gaussdev}), we use the convolution theorem and obtain
\begin{equation}
\label{eq:conv}
\hat{W}_{\delta}(w,k)=\hat{\delta}(k) \hat{\Psi}_g(w,k),
\end{equation}
where quantities with a hat are Fourier transforms of corresponding quantities. $\hat{\Psi}_g(w,k)$ is given by Eq.~(\ref{eq:gausswvft}), $\hat{\delta}(k)$ can be obtained by FFT from $\delta(x)$, and hence $W_{\delta}(w,x)$ can be obtained by the inverse FFT from Eq.~(\ref{eq:conv}).

Taking the Fourier transform of both sides of Eq.~(\ref{eq:invwv2}), we obtain
\begin{equation}
\label{eq:invwv2ft}
\hat{\delta}(k) = \int^{\infty}_{0}\hat{W}_{\delta}(w,k)\dd w.
\end{equation}
Notice that
\begin{equation}
\label{eq:wvftint}
\int^{\infty}_{0}\hat{\Psi}_g(w,k)\dd w=1,
\end{equation}
Eq.~(\ref{eq:invwv2ft}) can also be derived by integrating both sides of Eq.~(\ref{eq:conv}) with respect to $w$.

We see that the Fourier transform pair Eqs.~(\ref{eq:conv}) and (\ref{eq:invwv2ft}) are very simple and concise, and they compose the fast algorithm of wavelet transform.

\begin{figure}
\includegraphics[width=1.0\linewidth]{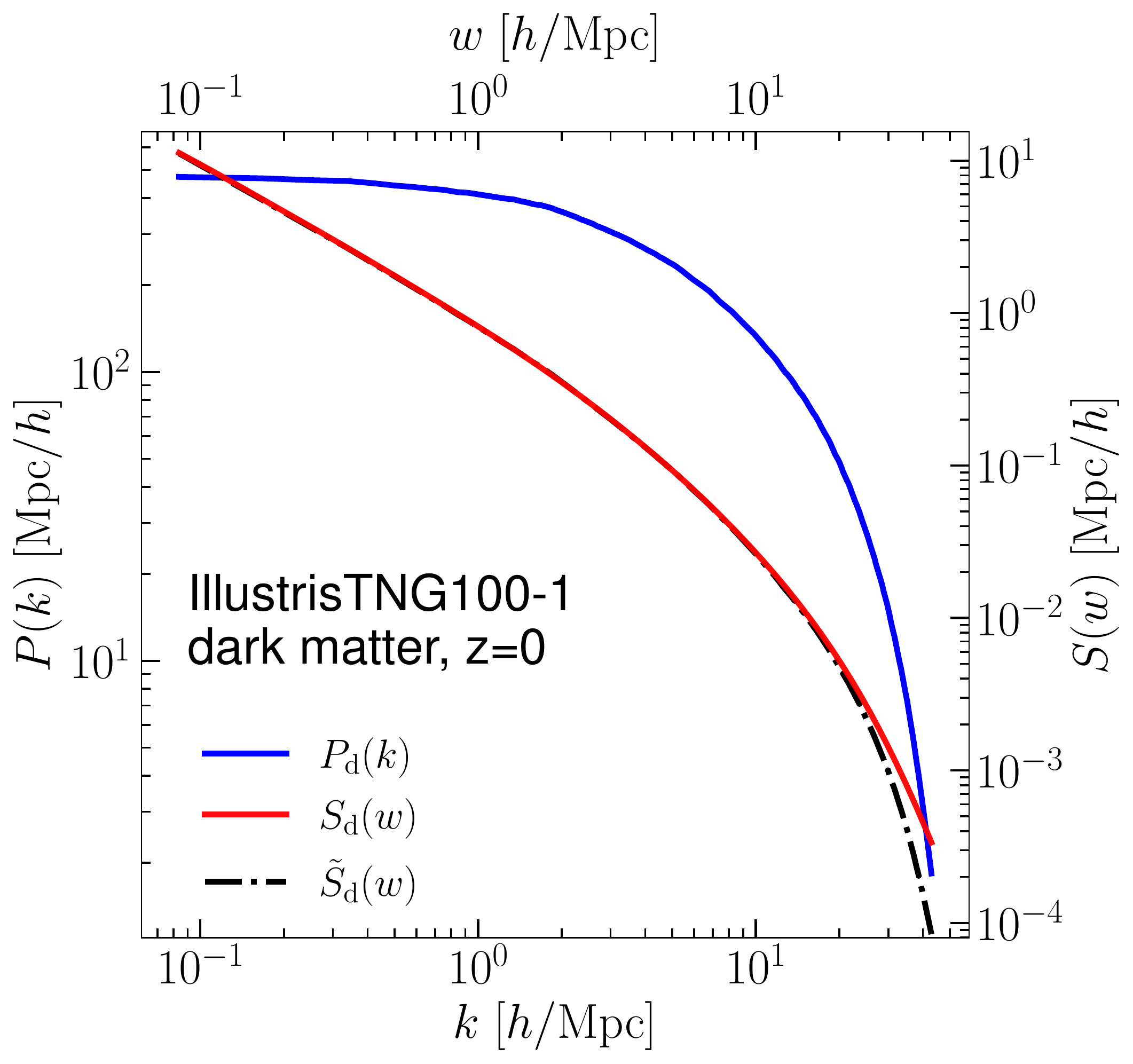}
\caption{The Fourier and wavelet power spectrum of the 1D density contrast field for dark matter. The left vertical axis is for the Fourier power spectrum, and the right axis is for the wavelet power spectrum. The data is taken from the IllustrisTNG simulation.}
\label{fig:dspm}
\end{figure}

\begin{figure}
\includegraphics[width=1.0\linewidth]{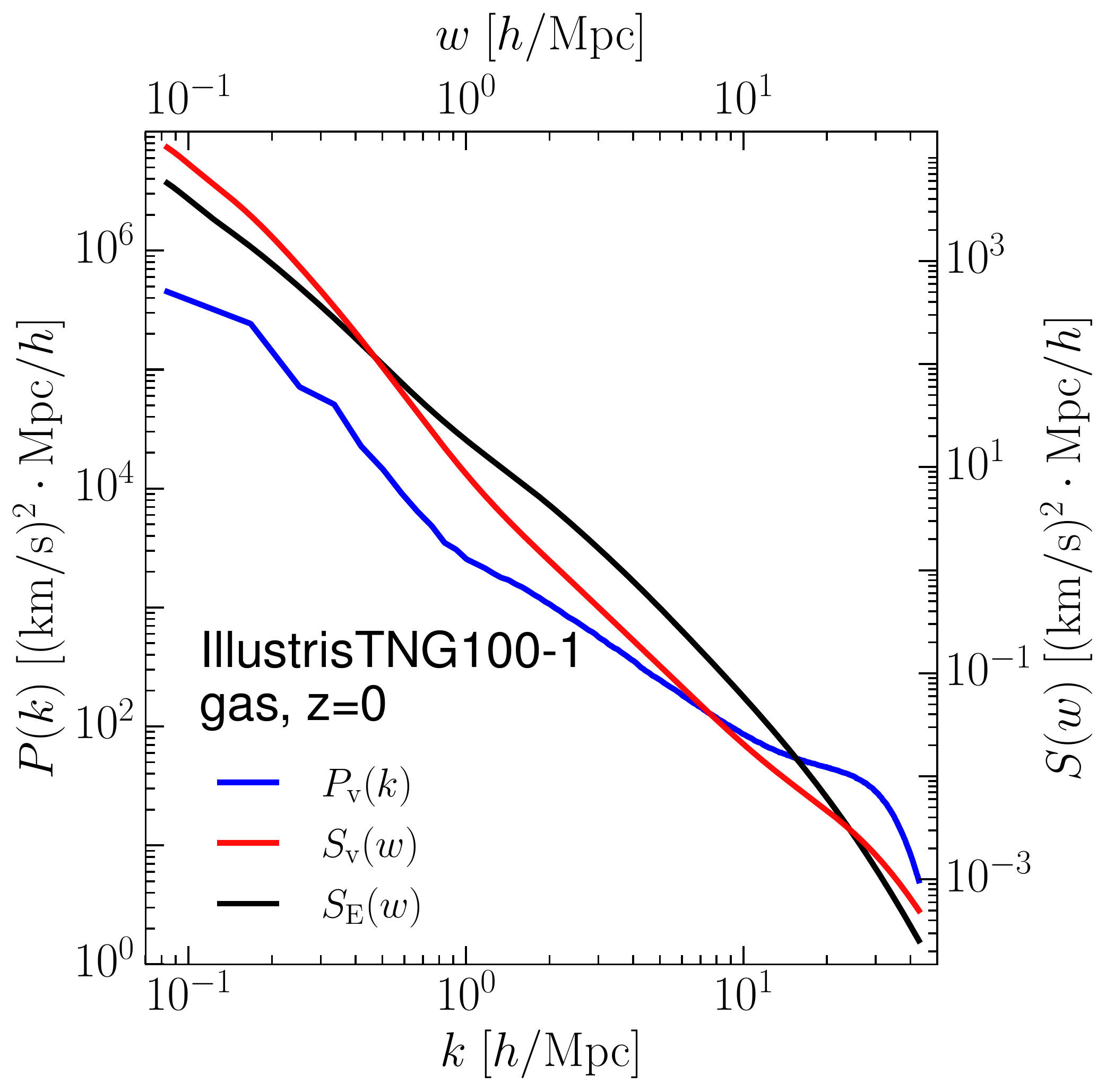}
\caption{The Fourier and wavelet power spectrum of the 1D velocity field for baryonic fluid. The left vertical axis is for the Fourier power spectrum, and the right axis is for the wavelet power spectrum. The wavelet energy spectrum is also shown for comparison. The data is taken from the IllustrisTNG simulation.}
\label{fig:vespm}
\end{figure}

\section{Applications in cosmology}
\label{sec:app}

We apply our Gaussian-derived wavelet to analyze the IllustrisTNG simulation data \cite{Nelson2019, Pillepich2018, Springel2018, Nelson2018, Naiman2018, Marinacci2018}, from which we select the sample {\tt IllustrisTNG100-1}, whose simulation box is $75{\rm Mpc}/h$ long. We use `cloud-in-cell' scheme \cite{Hockney1988} to assign all the particle (dark matter and baryonic) mass or velocity into a $1024^3$ mesh to acquire mass density or velocity at mesh points. Then we randomly select 100,000 lines, all vertical to the $x-y$ plane of the simulation box, and record all the relevant data at each line. In this way, we have 100,000 1D data. In Fig.~\ref{fig:denscalogram}, we show the dark matter density contrast of one such line data, together with its scalogram $W_{\delta}(w,x)$.

With the data, we compute the wavelet power spectrum of the density contrast field or peculiar velocity field. For one line data, the wavelet power spectrum can be defined as
\begin{equation}
\label{eq:wvlps}
S_i(w) \equiv \frac{1}{L^2_b}\int|W_i(w,x)|^2\dd x = \frac{1}{N_p L_b}\sum_{j=1}^{N_p}|W_i(w,x_j)|^2,
\end{equation}
where $i$ is the ID of the $i$-th line, $W_i(w,x)$ is the wavelet transform of the line data, $L_b=75{\rm Mpc}/h$ is the length of the simulation box, and $N_p=1024$ is the point number of the line data. Averaging over all the 100,000 lines, we obtain the total power spectrum as
\begin{equation}
\label{eq:wvtps}
S(w)=\frac{1}{N_l}\sum_{i=1}^{N_l}S_i(w),
\end{equation}
in which $N_l=100,000$.

We use Eq.~(\ref{eq:wvtps}) to compute the 1D wavelet power spectrum of the line data of dark matter density contrast, and show the power spectrum $S_{\rm d}(w)$ in Fig.~\ref{fig:dspm}, together with the 1D Fourier power spectrum $P_{\rm d}(k)$.

According to Parseval's theorem, we have
\begin{eqnarray}
\label{eq:parseval}
\int|W_{\delta}(w,x)|^2\dd x & = & \frac{1}{2\pi}\int|\hat{W}_{\delta}(w,k)|^2\dd k \nonumber \\
 & = & \frac{1}{2\pi} \int|\hat{\delta}(k)|^2 |\hat{\Psi}_g(w,k)|^2\dd k,
\end{eqnarray}
in which for the second equality we use Eq.~(\ref{eq:conv}). Eq.~(\ref{eq:parseval}) can provide an alternative method to compute the wavelet power spectrum in Eq.~(\ref{eq:wvlps}).

We can use Eq.~(\ref{eq:parseval}) to derive an interesting relationship between the Fourier and wavelet power spectrum. We assume the Fourier power spectrum $P_{\rm d}(k)$ scales as a power law of $k$, i.e. $P_{\rm d}(k)\sim|\hat{\delta}(k)|^2\sim k^{\alpha}$, in which $\alpha$ is the power index. If $\alpha>-5$, then from Eq.~(\ref{eq:parseval}) we have
\begin{eqnarray}
\label{eq:stdw}
\tilde{S}_{\rm d}(w) & \sim &\int|\hat{W}_{\delta}(w,x)|^2\dd x \nonumber \\
& \sim & \int P_{\rm d}(k) \hat{\Psi}_g(w,k)^2 \dd k \sim\frac{P_{\rm d}(w)}{w}.
\end{eqnarray}
From Fig.~\ref{fig:dspm}, we see that with a proper amplitude, the relationship by Eq.~(\ref{eq:stdw}) is quite accurate when $k<10h/{\rm Mpc}$.

With the wavelet transform $W_v(w,x)$ of the $z$-directional 1D velocity field for baryonic fluid, the wavelet power spectrum of the velocity field $S_{\rm v}(w)$, can be obtained in the same way as $S_{\rm d}(w)$ of the density contrast field. We show $S_{\rm v}(w)$ in Fig.~\ref{fig:vespm}, together with the Fourier power spectrum $P_{\rm v}(k)$ of the 1D velocity.

Turbulent flows in the baryonic fluid of the Universe are an important area in cosmological studies \cite{Hep2006, Zhu2010, Fang2011, Zhu2011, Zhu2013, Zhuravleva2014, Zhu2015, Yang2020}. The wavelet analysis technique is particularly suitable to investigate turbulent flows \cite{Farge1992}. In the scale (or Fourier) space, the Kolmogorov theory assumes the existence of an energy cascade between the different excited wavenumbers of the turbulent flow, while in the physical (or real) space, turbulent flows are characterized by complex multi-scale and chaotic motions, which can be classified into more elementary components, namely the coherent structures \footnote{ \url{https://en. wikipedia.org/ wiki/Coherent_turbulent_structure} }.

To study turbulence in the baryonic fluid of the Universe, which is distributed greatly inhomogeneously in space, the kinetic energy spectrum of the fluid is more desirable than its velocity spectrum. However, one cannot define the usual power spectrum for kinetic energy \cite{Bonazzola1987}. The reason can be seen from below
\begin{eqnarray}
\label{eq:kesp}
\frac{1}{2}\langle\rho(x)v^2(x)\rangle & = & \frac{1}{2}\int\rho(x)v^2(x)\dd x \nonumber \\
& = & \frac{1}{8\pi^2}\int\hat{\rho}^{*}(k_1+k_2)\hat{v}(k_1)\hat{v}(k_2)\dd k_1 \dd k_2 \nonumber \\
& = & \frac{1}{8\pi^2}\int B_{\rho v}(k_1, k_2)\dd k_1 \dd k_2,
\end{eqnarray}
where we do not have a power spectrum and we have to define a mixed bispectrum as $B_{\rho v}(k_1, k_2)\equiv \hat{\rho}^{*}(k_1+k_2) \hat{v}(k_1) \hat{v}(k_2)$ \cite{Bernardeau2002}. With the wavelet transform, however, we can define a kinetic energy power spectrum for a 1D data readily as below
\begin{equation}
\label{eq:wvesp}
S_{{\rm E},i}(w) = \frac{1}{2L^2_b}\int\Delta_i(x)|W_{v,i}(w,x)|^2\dd x,
\end{equation}
in which $\Delta(x)=\rho_b(x)/\bar{\rho}_b$ is the dimensionless baryonic density. Averaging over the total 100,000 lines by using Eq.~(\ref{eq:wvtps}), we can obtain the wavelet kinetic energy power spectrum $S_{\rm E}(w)$, which is also shown in Fig.~\ref{fig:vespm}. From another viewpoint, one can consider $S_{\rm E}(w)$ as the density-weighted velocity power spectrum. Moreover, due to the localization property of the wavelet transform, $S_{\rm E}(w)$ can be generalized to the spectrum with the space-restricted (e.g. $x_1<x<x_2$) or the density-restricted (e.g. $\rho_1<\rho<\rho_2$) integration, which may be more suitable for investigations of inhomogeneous turbulent flows.

\begin{figure}
\includegraphics[width=1.0\linewidth]{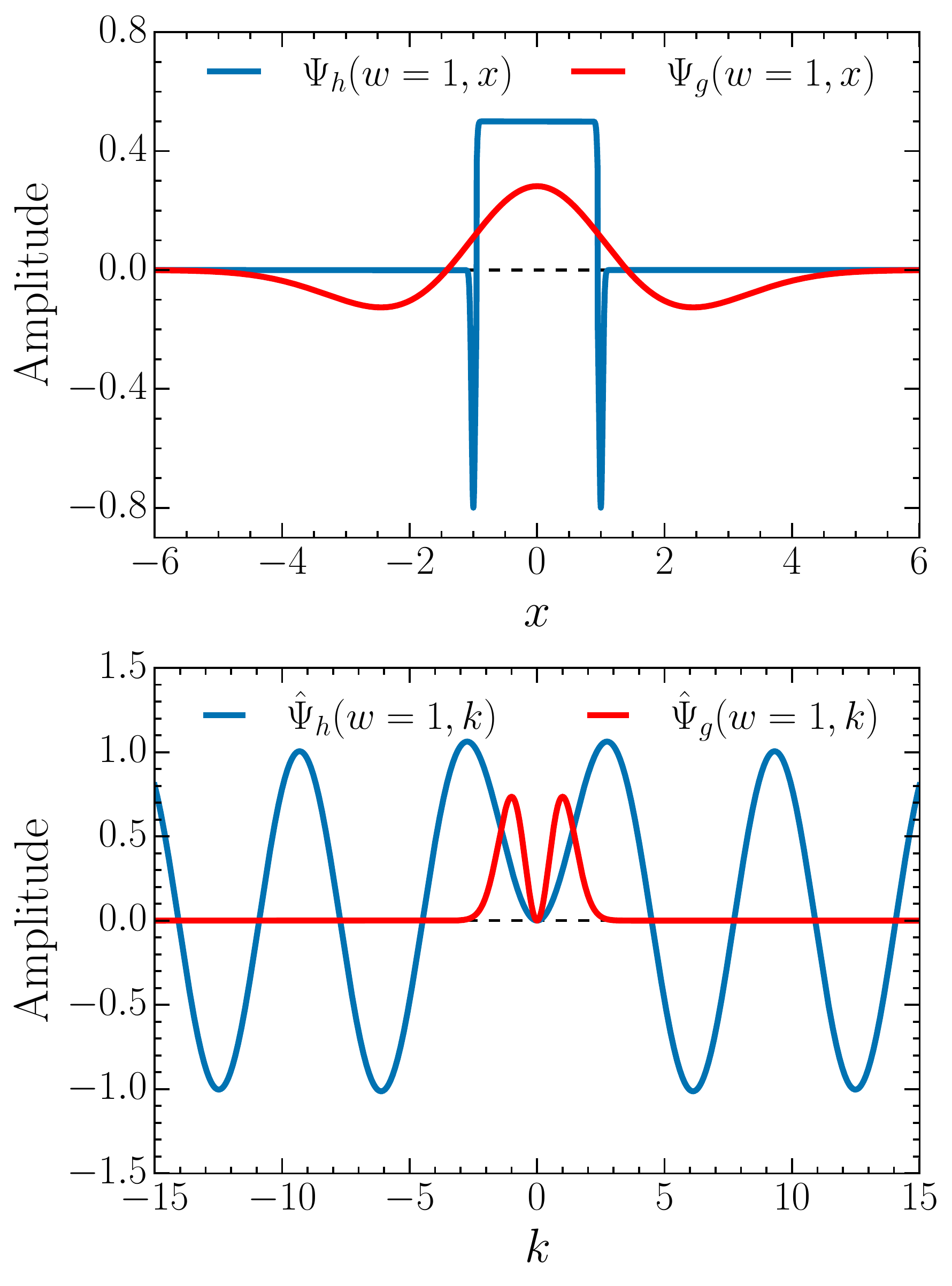}
\caption{The functional form of the counter-example (upper panel), and its Fourier transform (lower panel) in Section~\ref{sec:disc}, shown as blue lines. The Gaussian-derived wavelet and its Fourier transform are also shown for comparison (red lines).}
\label{fig:no_go}
\end{figure}

\section{Discussions}
\label{sec:disc}

Our wavelet scheme is a general method, which can be used to construct a large category of CWTs, but we emphasize that Eqs.~(\ref{eq:gausswv}) and (\ref{eq:wvft2}) are not sufficient conditions to construct a continuous wavelet. We show this with a counter-example in the following. The top-hat function in real space with the scale parameter $w$,
\begin{eqnarray}
\label{eq:toph}
h_w(x)=\left\{\begin{matrix}
 & \frac{w}{2}, & -\frac{1}{w} \leq x \leq \frac{1}{w}, \\ \\
 & 0,           & {\rm otherwise},
\end{matrix}\right.
\end{eqnarray}
is a smoothing function, whose Fourier transform is $\hat{h}_w(k) = (w/k)\sin(k/w)$. According to our scheme, however, its derivative with respect to $w$,
\begin{equation}
\label{eq:kthfun}
\hat{\Psi}_h(w,k) \equiv \frac{\partial\hat{h}_w(k)}{\partial w} = \frac{1}{k}\sin\left(\frac{k}{w}\right) - \frac{1}{w}\cos\left(\frac{k}{w}\right),
\end{equation}
is NOT a continuous wavelet, since $\hat{\Psi}_h(w,k)$ is oscillatory when $k\rightarrow\infty$, and hence does not satisfy the square-integrable condition Eq.~(\ref{eq:sicond}). The real space couterpart of Eq.~(\ref{eq:kthfun}) is
\begin{eqnarray}
\label{eq:rthfun}
\Psi_h(w,x) & \equiv & \frac{\partial h_w(x)}{\partial w} =  -\frac{1}{2w}[\delta_D(x+\frac{1}{w})+\delta_D(x-\frac{1}{w})] \nonumber \\ &  + &  \begin{cases}
  \frac{1}{2}, & -\frac{1}{w} < x < \frac{1}{w}, \\
    0,          & \mathrm{otherwise}.
 \end{cases}
\end{eqnarray}
To facilitate understanding of this counter-example, we show $\Psi_h(w,x)$ and $\hat\Psi_h(w,k)$ in Fig.~\ref{fig:no_go}.

Generally, since the smoothing function $s_w(x)$ is an even function, its Fourier transform can be formally expressed as a Taylor expansion as follows,
\begin{equation}
\label{eq:sft}
\hat{s}_w(k) = 1 + \frac{\hat{s}^{(2)}(0)}{2!}\left(\frac{k}{w}\right)^2 + \frac{\hat{s}^{(4)}(0)}{4!}\left(\frac{k}{w}\right)^4 + ...,
\end{equation}
from which we know $\hat{s}_w(0)=1$, consistent with the normalization condition Eq.~(\ref{eq:norm}). From Eq.~(\ref{eq:sft}), we can define a function $\Psi_s(w,x)$ by its Fourier transform as
\begin{equation}
\label{eq:swv}
\hat{\Psi}_s(w,k)\equiv\frac{\partial\hat{s}_w(k)}{\partial w} = -\frac{k}{w}\frac{\partial\hat{s}_w(k)}{\partial k}.
\end{equation}
It is easy to see that $\hat{\Psi}_s(w,k)$ satisfies the admissibility condition of wavelet, i.e. $\hat{\Psi}_s(w,0)=0$. If it also satisfies the square-integrable condition Eq.~(\ref{eq:sicond}), then $\Psi_s(w,x)$ should be a continuous wavelet.

Additionally, it is not difficult to generalize our scheme to the three-dimensional (3D) case. For example, the 3D anisotropic Gaussian function is
\begin{equation}
\label{eq:gassian3d}
g_{\boldsymbol{w}}(\boldsymbol{x}) = \prod_{i=1}^{3} g_{w_i}(x_i) = \frac{w_1w_2w_3}{8\pi^{3/2}} e^{-\frac{1}{4}(w_1^2x_1^2 + w_2^2x_2^2 + w_3^2x_3^2)},
\end{equation}
where $\boldsymbol{w}=(w_1, w_2, w_3)$, $\boldsymbol{x}=(x_1, x_2, x_3)$. Hence the 3D anisotropic Gaussian-derived wavelet can be obtained as
\begin{eqnarray}
\label{eq:3dgaussianwv}
\Psi_{g}({\boldsymbol w}, {\boldsymbol x}) & \equiv & \frac{\partial^3g_{\boldsymbol w}({\boldsymbol x})}{\partial w_1 \partial w_2 \partial w_3} \nonumber \\
 & = & \prod_{i=1}^{3} \frac{\partial g_{w_i}(x_i)}{\partial w_i} = \prod_{i=1}^{3} \Psi_{g}(w_i, x_i).
\end{eqnarray}
We see that the 3D anisotropic wavelet is not a 3D Mexican hat wavelet. In the future, we will explore the possible applications of this 3D wavelet in cosmology.

\section{Summary and Conclusions}
\label{sec:conc}

The discrete wavelet transforms are constructed by dilation and translation both dyadically in the scale and position. They can provide a set of complete and orthonormal bases, based on which fast algorithms can be designed, and hence DWTs are very suitable for technical applications, such as data compactness, or image processing. While in some areas, such as cosmological investigations, the localization of continuous wavelets in both scale and position is much more desirable than the orthogonality of discrete wavelets. Nevertheless, the usual CWTs are not convenient to use since the inverse wavelet transforms are two-dimensional integrations, which may be computationally very cumbersome and time-consuming.

In this work, we present a novel scheme of constructing continuous wavelet functions, in which the wavelet functions are obtained by taking the first derivative of smoothing functions with respect to the scale parameter. Due to this wavelet constructing scheme, the inverse transforms are only 1D integrations with respect to the scale parameter, and hence CWTs are more ready to use than the usual scheme.

We then apply the Gaussian-derived wavelet constructed by our scheme to computations of the density power spectrum for dark matter, the velocity power spectrum and the kinetic energy spectrum for baryonic fluid. These computations exhibit the convenience and strength of the CWTs. From the transform pairs Eqs.~(\ref{eq:gaussdev}) and (\ref{eq:invwv2}) in real space, and Eqs.~(\ref{eq:conv}) and (\ref{eq:invwv2ft}) in Fourier space, we see that our wavelet transform scheme is very simple, and we believe that the simplicity of our scheme will make CWTs very useful in cosmology.

\section*{Acknowledgements}

We acknowledge the support by the National Science Foundation of China (No. 11947415, 12047569), and by the Natural Science Foundation of Jilin Province, China (No. 20180101228JC). In this work, we used the data from IllustrisTNG simulations. The IllustrisTNG simulations were undertaken with compute time awarded by the Gauss Centre for Supercomputing (GCS) under GCS Large-Scale Projects GCS-ILLU and GCS-DWAR on the GCS share of the supercomputer Hazel Hen at the High Performance Computing Center Stuttgart (HLRS), as well as on the machines of the Max Planck Computing and Data Facility (MPCDF) in Garching, Germany.


\begin{appendices}
\numberwithin{equation}{section}
\numberwithin{figure}{section}

\section{Derivation of the traditional continuous wavelet transform} \label{appendix:traditional_cwt}

Given a wavelet function $\psi(x)$, the traditional CWT of a function $f(x)$ is defined as the convolution of $f(x)$ with a scaled version of $\psi(x)$. Or to put it another way, the traditional CWT can be viewed as the projection of $f(x)$ on the dilation and translation of $\Psi(x)$ \cite{Mallat1999}. The traditional CWT is shown in Eq.~(\ref{eq:cwt}), and again as follows
\begin{equation}
\label{eq:cwt_appendix}
WT_f(a,b) = \frac{1}{\sqrt{|a|}}\int^{\infty}_{-\infty} f(x)\bar{\psi}(\frac{x-b}{a})\mathrm{d}x.
\end{equation}
With Parseval's theorem for traditional CWT (see theorem 3.10 in \cite{Chui1992} for its proof)
\begin{align}
\label{eq:parseval_cwt}
C_\psi& \int_{-\infty}^{+\infty} f(x)\bar g(x)\mathrm{d}x \nonumber \\
  & = \int_{-\infty}^{+\infty}\int_{-\infty}^{+\infty} WT_f(a,b)\overline{WT}_g(a,b)\frac{\mathrm{d}a\mathrm{d}b}{a^2},
\end{align}
where $C_\psi$ is given in Eq.~(\ref{eq:cpsi}). Let $g(x)$ be the Dirac delta function, i.e. $g(x)=\delta_D(x-x')$, and from Eq.~(\ref{eq:parseval_cwt}) we obtain
\begin{equation}
\label{eq:icwt_appendix}
f(x) = C^{-1}_{\psi}\int^{\infty}_{-\infty}\int^{\infty}_{-\infty} WT_f(a,b) \frac{1}{\sqrt{|a|}}\psi(\frac{x-b}{a})\frac{\mathrm{d}a\mathrm{d}b}{a^2},
\end{equation}
which is the reconstruction formula or inverse transform of the conventional method. Notice that, although both traditional CWT Eq.~(\ref{eq:cwt_appendix}) and our CWT Eq.~(\ref{eq:gaussdev}) are the convolution of a general function with a wavelet function, the traditional reconstruction formula Eq.~(\ref{eq:icwt_appendix}) is far more complex than our method Eq.~(\ref{eq:invwv2}).

It should be pointed out that the traditional CWT does not tell us how to construct a specific wavelet function. In fact, each existing continuous wavelet is constructed in some unique way. For example, the well-known Mexican hat wavelet is defined as the negative normalized second derivative of the Gaussian function, i.e.
\begin{equation} 
\label{eq:def_mexican_hat}
\Psi_\mathrm{MH}(x)\equiv C\frac{\mathrm{d}^2}{\mathrm{d}x^2}\Big[ \frac{1}{\sigma \sqrt{2\pi}}e^{-\frac{x^2}{2\sigma^2}}\Big],
\end{equation}
where $C$ is a normalization constant, such that $\int_{-\infty}^{+\infty}|\psi(x)|^2\mathrm{d}x=1$. If the standard deviation $\sigma$ is set to $\sqrt{2}$, then we have
\begin{equation}
\label{eq:mexican_hat}
\Psi_\mathrm{MH}(x)=\frac{1}{(2\pi)^{1/4}\sqrt{3}}(2-x^2)e^{-\frac{x^2}{4}},
\end{equation}
which is the same as our wavelet Eq.~(\ref{eq:gausswv}), except for the prefactor. However, the original function cannot be reconstructed by integrating $WT_f(a,b)$ with respect to the scale parameter $a$ directly. In order to better understand this point, we perform the Fourier transform of Eq.~(\ref{eq:cwt_appendix}) as
\begin{equation}
\label{eq:traditional_cwt_k}
\widehat{WT}_f(a,k)=\hat f(k)\frac{a}{\sqrt{|a|}}\hat\Psi_\mathrm{MH}(ak),
\end{equation}
where $\hat\Psi_\mathrm{MH}(k)=4(8\pi/9)^{1/4}k^2e^{-k^2}$. Obviously, due to $\int_{-\infty}^{+\infty} (a/\sqrt{|a|}) \hat\Psi_\mathrm{MH}(ak) \mathrm{d}a=0$, we cannot recover the original function $\hat f(k)$ with $\int_{-\infty}^{+\infty} \widehat{WT}_f(a,k)\mathrm{d}a$.

As another example, we describe briefly how to design the Meyer wavelet (see \cite{Mallat1999} for details). The Meyer wavelet is constructed by means of the multiresolution analysis in the Fourier space, which is much more tedious than constructing the Mexican hat wavelet. The procedure consists of two steps, i.e. constructing the scaling function with the conjugate mirror filter and then constructing the wavelet function with the wavelet equation. The conjugate mirror filter $\hat h(k)$ used here is given by
\begin{equation} 
\label{eq:filters}
\hat h(k)=
\begin{cases}
  \sqrt{2}, & |k|\leqslant \frac{\pi}{3}, \\
  \sqrt{2}\cos\left[ \frac{\pi}{2}\beta(\frac{3|k|}{\pi}-1)\right], & \frac{\pi}{3}< |k|\leq \frac{2\pi}{3},\\
  0, & \frac{2\pi}{3}<|k|\leqslant \pi,
\end{cases}
\end{equation}
where $\beta(t)$ is an auxiliary function such that $|\hat h(k)|^2+|\hat h(k+\pi)|^2=2$, which indicates the scaling functions are orthonormal under integer translations. The simplest case of $\beta(t)$ \cite{Vermehren2015} is
\begin{equation} 
\label{eq:aux_func}
\beta(x)=\begin{cases}
	0, & x\leqslant 0,\\
	x, & 0<x\leqslant 1, \\
	1, & x>1.
	\end{cases}
\end{equation}
Cutting off the terms of $p>1$ in the scaling function $\hat\phi(k)=\prod_{p=1}^{+\infty}\hat h(2^{-p}k)/\sqrt{2}$ and letting $\hat\phi(k)=0$ for $|k|> 4\pi/3$, we have
\begin{equation} 
\label{eq:meyer_scaling_function}
\hat\phi(k)=
    \begin{cases}
      1, & |k|\leqslant \frac{2\pi}{3}, \\
      \sin\left( \frac{3|k|}{4}\right), & \frac{2\pi}{3}< |k|\leqslant \frac{4\pi}{3}, \\
      0, & |k|> \frac{4\pi}{3},
    \end{cases}
\end{equation}
which is the Meyer scaling function. Then substituting Eq.~(\ref{eq:filters}) and Eq.~(\ref{eq:meyer_scaling_function}) into the wavelet equation, $\hat\psi(k) = \hat g(k/2) \hat\phi(k/2)/\sqrt{2}$, where $\hat g(k)=\mathrm{e}^{-ik}\bar{\hat h}(k+\pi)$, we obtain the Meyer wavelet function as below,
\begin{equation} 
\label{eq:meyer_wavelet}
\hat{\Psi}_\mathrm{MY}\left( k \right) \equiv
   \begin{cases}
       -\cos \left( \frac{3|k|}{4}\right) e^{-ik/2},&		2\pi /3<|k|\leqslant 4\pi /3,\\
       \sin \left( \frac{3|k|}{8}\right) e^{-ik/2},&		4\pi /3<|k|\leqslant 8\pi /3,\\
       0, &		\text{otherwise},\\
   \end{cases}
\end{equation}
which is a complex function and incompatible with our scheme.

Unlike conventional continuous wavelet functions designed in various ways, our scheme offers a uniform approach to construct wavelets by differentiating a smoothing function with respect to its scale parameter $w$. Because of this wavelet-constructing method, the original signal can be reconstructed easily by integrating the CWT $W(w,x)$ with respect to $w$, and hence, our method is much simpler than the traditional method.

\section{Comparison with wavelets derived from different smoothing functions}

\begin{figure*}
\centering
\includegraphics[width=0.90\linewidth]{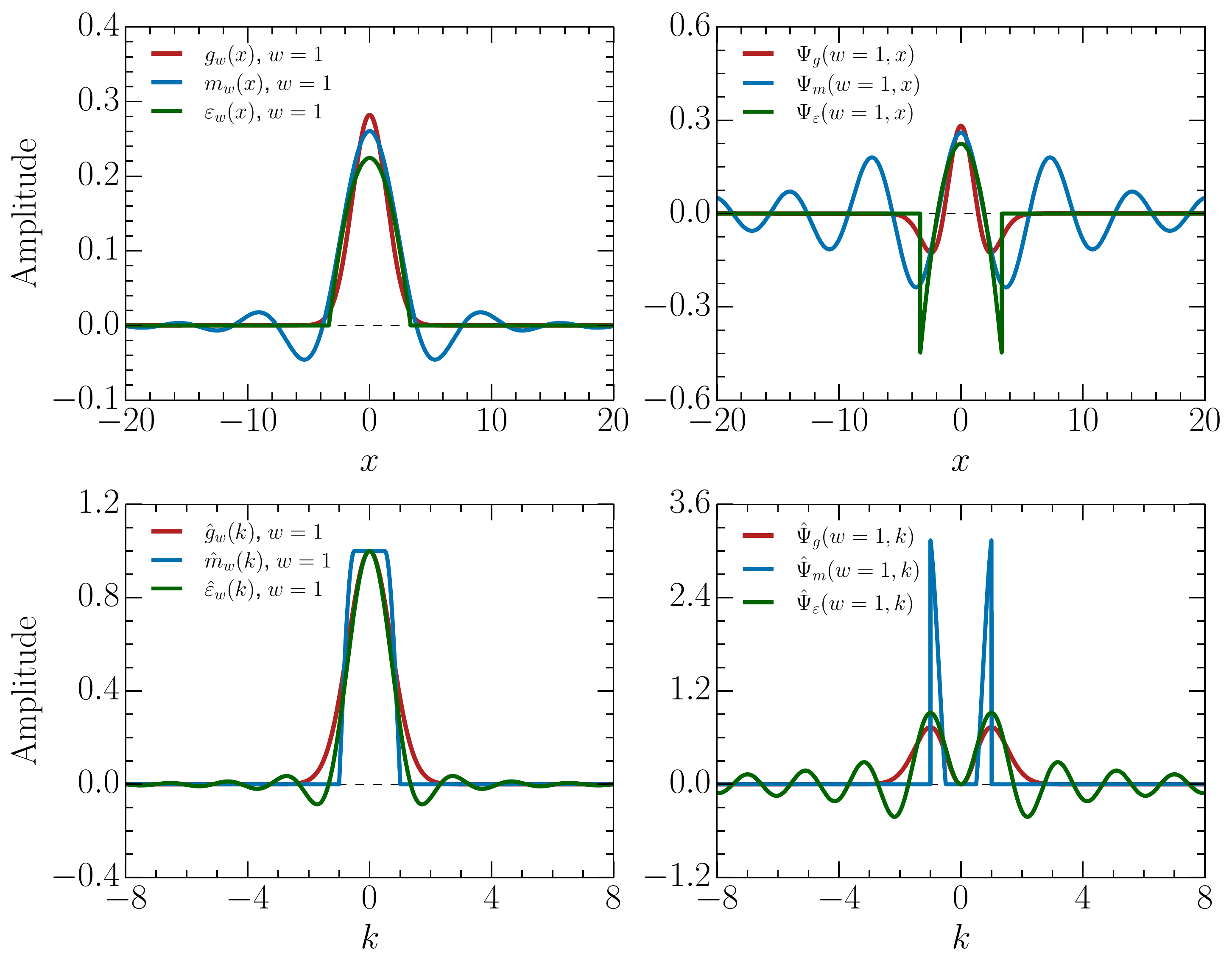}
\caption{Three different smoothing functions and their corresponding wavelets. Upper left panel: Gaussian (red line), Meyer scaling (blue line) and Epanechnikov (green line) function in real space. Upper right panel: the wavelet functions derived from these smoothing functions. Lower left panel: Fourier transforms of the three smoothing functions. Lower right panel: Fourier transforms of the wavelets derived from the corresponding smoothing functions.}
\label{fig:three_wavelets}
\end{figure*}

In this work, we use the Gaussian function as the smoothing function to derive the wavelet function, due to its simplicity and wide applications. The Gaussian-derived wavelet is localized in both real and Fourier space. In other words, both $\Psi_g(w,x)$ and $\hat \Psi_g(w,k)$ are approximately compactly supported, as shown in Fig.~\ref{fig:wvlet}. Besides the Gaussian function, we also consider other bell-shaped functions, such as the Meyer scaling function \cite{Vermehren2015} and the Epanechnikov function \cite{Epanechnikov1969}, as the smoothing function. The expressions of Meyer scaling function and Epanechnikov function are
\begin{equation} 
\label{eq:meyer_scaling}
m_w\left( x \right) =\begin{cases}
	\left( \frac{1}{2\pi}+\frac{1}{\pi ^2} \right) w, &		x=0, \\ \\
	\frac{\pi \sin \left( wx/2 \right) + wx\cos \left( wx \right)}{\pi ^2x-w^2x^3}, &		x\ne 0, \\
  \end{cases}
\end{equation}
and
\begin{equation} 
\label{eq:epan_func}
\varepsilon _w\left( x \right) =\begin{cases}
\frac{3w}{4A}\left( 1-\left( \frac{wx}{A} \right) ^2 \right), &		-\frac{A}{w}\leqslant x\leqslant \frac{A}{w}, \\ \\
	  0, &		\mathrm{otherwise},
      \end{cases}
\end{equation}
respectively, where the constant $A$ is approximately equal to $3.342$. We will explain the reason of such a choice below. The Fourier transforms of Eq.~(\ref{eq:meyer_scaling}) and Eq.~(\ref{eq:epan_func}) are given by
\begin{equation} 
\label{eq:meyer_scaling_k}
\hat{m}_w\left( k \right)  =
  \begin{cases}
	  1, &		|k|\leqslant \frac{w}{2},\\
	  \sin \left( \frac{\pi |k|}{w} \right), &		\frac{w}{2}<|k|\leqslant w,\\
	  0, &		\text{otherwise}, \\
  \end{cases}
\end{equation}
and
\begin{equation} 
\label{eq:epan_func_k}
  \hat{\varepsilon}_w\left( k \right) =
  \begin{cases}
	  1, &    k=0,  \\ \\
 	 \frac{-(3 Ak/w)\cos(Ak/w) + 3\sin(Ak/w)}{(Ak/w)^3},    &   k \ne 0.
  \end{cases}
\end{equation}
The corresponding shapes of these smoothing functions in real space are shown in the upper left panel of Fig.~\ref{fig:three_wavelets}, and the lower left panel is for their Fourier transforms.

Next, according to our scheme, we define the wavelets as the first derivative of smoothing functions with respect to $w$. Hence the Meyer-derived wavelet in real space is
\begin{align} 
\label{eq:meyer_derived_w}
  \Psi _m\left( w,x \right) \equiv \frac{\partial m_w(x)}{\partial w}=\begin{cases}
 	 -\left( \frac{3}{16}+\frac{1}{4\pi ^2} \right), &		x=\pm \frac{\pi}{w}, \\ \\
	 \Psi_0(w,x), &		x\ne \pm \frac{\pi}{w},
  \end{cases}
\end{align}
where the form of $\Psi_0(w,x)$ is
\begin{align}
      \Psi_0(w,x) &=\frac{\cos \left( wx \right) -\sin \left( \frac{wx}{2} \right)}{2\left( wx+\pi \right) ^2}+\frac{\sin \left( \frac{wx}{2} \right) +\cos \left( wx \right)}{2\left( wx-\pi \right) ^2}\nonumber		\\
	& +\frac{2\sin \left( wx \right) -\cos \left( \frac{wx}{2} \right)}{4\left( wx-\pi \right)} +\frac{2\sin \left( wx \right) +\cos \left( \frac{wx}{2} \right)}{4\left(wx + \pi \right)}.
\end{align}
The Epanechnikov-derived wavelet function is
\begin{align} 
\label{eq:epan_derived_w}
  \Psi _{\varepsilon}\left( w,x \right) &\equiv \frac{\partial \varepsilon_w(x)}{\partial w} \nonumber \\ &=\begin{cases}
  \frac{3}{4A}\left[ 1-3\left( \frac{wx}{A} \right) ^2 \right] ,&		-\frac{A}{w}\leqslant x\leqslant \frac{A}{w}, \\ \\
	  0, &		\text{otherwise}.
  \end{cases}
\end{align}
The Fourier transforms of Eq.~(\ref{eq:meyer_derived_w}) and Eq.~(\ref{eq:epan_derived_w}) are given as
\begin{equation} 
\label{eq:meyer_derived_wk}
   \hat{\Psi}_m\left( w,k \right) =\begin{cases}
   -\frac{\pi \left| k \right|\cos \left( \frac{\pi \left| k \right|}{w} \right)}{w^2},&		\frac{w}{2}<|k|\leqslant w,  \\ \\
	0, &		\text{otherwise},
  \end{cases}
\end{equation}
and
\begin{equation} 
\label{eq:epan_derived_wk}
   \hat{\Psi}_{\varepsilon}\left( w,k \right) =\begin{cases}
	0,&		k=0, \\ \\
	-\frac{3\left[ \left( \frac{Ak}{w} \right) ^2-3 \right] \sin \left( \frac{Ak}{w} \right) +9
    \left( \frac{Ak}{w} \right) \cos \left( \frac{Ak}{w} \right)}{w\left( \frac{Ak}{w} \right) ^3}, &		k\ne 0,
   \end{cases}
\end{equation}
respectively. It is easy to verify that both the Meyer- and the Epanechnikov-derived wavelet satisfy the admissibility condition and the square-integrable condition. Notice that the scale parameter $w$ from Eq.~(\ref{eq:meyer_scaling}) to Eq.~(\ref{eq:epan_derived_wk}) is defined as the frequency at which the first peak of $\hat\Psi(w,k)$ is located when $k/w=\pm1$, i.e. the frequency at which $\hat\Psi(w,k)$ takes the maximum value when $k/w=\pm1$, as shown in the lower right panel in Fig.~\ref{fig:three_wavelets}. That is the reason we choose $A=3.342$ for Epanechnikov-derived wavelet.

With these two wavelets, we can now compare their properties with those of the Gaussian-derived wavelet. By visual inspection of the upper right panel of Fig.~\ref{fig:three_wavelets}, we find that when $x$ is farther away from the origin in real space, the Meyer-derived wavelet is still oscillatory obviously while other wavelets are already close to zero. On the other hand, as shown in the lower right panel of Fig.~\ref{fig:three_wavelets}, the Epanechnikov-derived wavelet is more extended in the Fourier space than others. Hence, compared with the Gaussian-derived wavelet, the Meyer- and the Epanechnikov-derived wavelet are not well localized in both real and Fourier space simultaneously, which degrades their performance and applicability in practice.

\section{Comparing the power spectrum computed with the traditional continuous wavelet transform}
\begin{figure}
\includegraphics[width=0.90\linewidth]{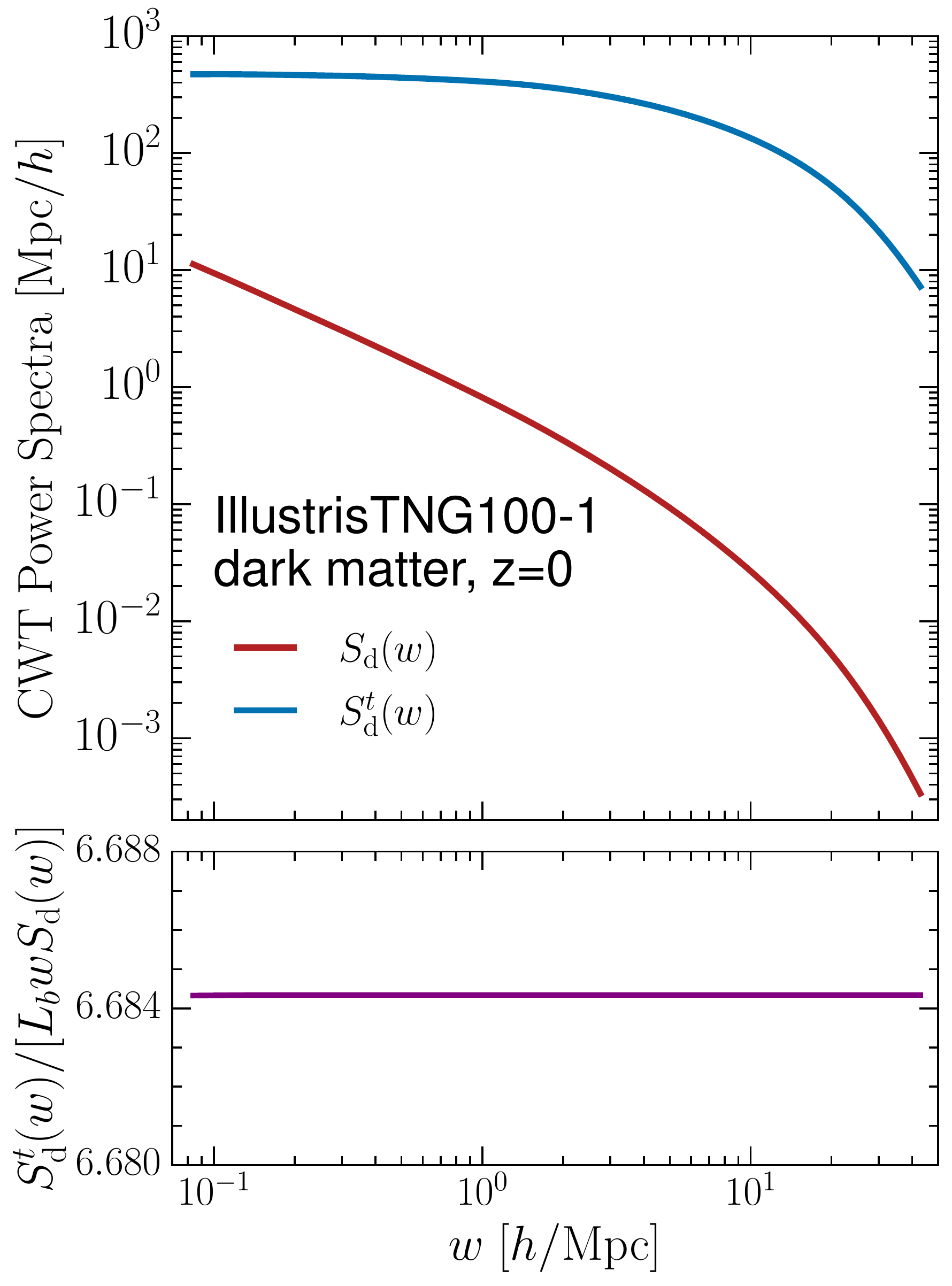}
\caption{The relationship between our wavelet power spectrum and the traditional CWT power spectrum. Upper panel: the Gaussian-derived wavelet power spectrum $S_\mathrm{d}(w)$ (red line), and the traditional wavelet power spectrum $S^t_\mathrm{d}(a=1/w)$ (blue line). Lower panel: the ratio of $S^t_\mathrm{d}(w)$ to $L_bwS_\mathrm{d}(w)$.}
\label{fig:traditionalCWT}
\end{figure}

In order to fully demonstrate the strength and usefulness of our scheme, we compute the 1D power spectrum for the same data set of the dark matter density field at $z=0$, with the traditional CWT and our CWT, respectively. Due to its similarity to the Gaussian-derived wavelet, we use the Mexican hat wavelet (shown in Eq.~(\ref{eq:mexican_hat})) as the basis function of the traditional wavelet transform for the computation. Here, the wavelet power spectrum averaged for 100,000 lines is defined as
\begin{equation} 
\label{eq:tcwt_power}
  S^t_\mathrm{d}(a)=\frac{1}{N_l}\sum_{i=1}^{N_l}\Big\{\frac{1}{L_b}\int WT_\delta(a,b)^2\mathrm{d}b \Big\}_i,
\end{equation}
where $N_l=100,000$, and the superscript `$t$' denotes `traditional' to distinguish from our CWT power spectrum. Letting $w=1/a$, we put $S^t_\mathrm{d}(a)$ and $S_\mathrm{d}(w)$ together in Fig.~\ref{fig:traditionalCWT}, and find that the magnitude of $S^t_\mathrm{d}(w)$ is greater than that of $S_\mathrm{d}(w)$ in the entire scale range.

We reveal the relationship between these two wavelet power spectra in the following. Combining Eqs.~(\ref{eq:cwt_appendix}), (\ref{eq:mexican_hat}) and (\ref{eq:tcwt_power}), and with $w = 1/a$, we obtain
\begin{equation}
\label{eq:tcwt_power_a}
S^t_\mathrm{d}(w)=\frac{w}{3\sqrt{2\pi}L_b} \frac{1}{N_l}\sum_{i=1}^{N_l}\Big\{\int I(w,b)\mathrm{d}b \Big\}_i,
\end{equation}
in which $I(w,b)$ is defined as
\begin{equation}
\label{eq:Ifunc}
I(w,b)\equiv\iint \mathrm{d}x\mathrm{d}x'\delta(x)\delta(x')\Psi_I[w(x-b)]\Psi_I[w(x'-b)],
\end{equation}
where $\Psi_I(x)=(2-x^2)e^{-x^2/4}$. Substituting Eq.~(\ref{eq:gaussdev}) into Eq.~(\ref{eq:wvtps}), we get
\begin{equation}
\label{eq:cwt_power_a}
S_\mathrm{d}(w)=\frac{1}{16\pi L_b^2}\frac{1}{N_l}\sum_{i=1}^{N_l}\Big\{\int I(w,b)\mathrm{d}b \Big\}_i.
\end{equation}
From Eqs.~(\ref{eq:tcwt_power_a}) and (\ref{eq:cwt_power_a}), one can see that $S^t_\mathrm{d}(w)$ and $S_\mathrm{d}(w)$ satisfy
\begin{equation}
\label{eq:relation_two_power}
\frac{S^t_\mathrm{d}(w)}{L_bwS_\mathrm{d}(w)} = \frac{16\pi }{3\sqrt{2\pi}}\approx 6.6843,
\end{equation}
which is reproduced by our numerical result shown in the lower panel of Fig.~\ref{fig:traditionalCWT}. Eq.~\ref{eq:relation_two_power} indicates that our CWT and the traditional CWT are actually equivalent to each other in applicability.

\end{appendices}

\end{document}